\documentclass[twocolumn]{aa}

\usepackage{amsmath}
\usepackage[varg]{txfonts}
\usepackage{graphicx}
\usepackage{epstopdf}
\usepackage{hyperref}

\title{Second generation planet formation in NN\,Serpentis?}
\titlerunning{Second Generation Planet Formation in NN\,Ser}
\author{M. V\"olschow \inst{\ref{inst1}}
\and R. Banerjee \inst{\ref{inst1}}
\and F. V. Hessman \inst{\ref{inst2}} }
\authorrunning{V\"olschow et al.}
\institute{Hamburger Sternwarte, Universit\"at Hamburg, Gojenbergsweg
  112, 21029 Hamburg, Germany \label{inst1}
  \and
  Institut f\"ur Astrophysik, Universit\"at G\"ottingen, Friedrich-Hund-Platz 1,
37077 G\"ottingen, Germany \label{inst2}}
\offprints{{\tt marcel.voelschow@uni-hamburg.de}}
\date{}

\newcommand{\simge} {\,{}^>_{\sim}\,}

\def\D{\mathrm{d}}

\newcommand{\cm}{\rm{cm}}

\newcommand{\km}{\rm{km}}
\newcommand{\AU}{\rm{au}}
\newcommand{\s}{\rm{s}}

\newcommand{\g}{\rm{ g}}

\newcommand{\Msol}{{\rm ~M}_{\sun}}
\newcommand{\Mjup}{{\rm ~M}_{Jup}}

\newcommand{\au}{\rm{au}}

\newcommand{\yr}{\rm{yr}}

\newcommand{\pcc}{{\rm ~cm}^{-3}}

\newcommand{\bef}{\begin{figure}[!t]}
\newcommand{\eef}{\end{figure}}

\abstract{
In this paper, we study the general impact of stellar mass-ejection events in planetary orbits in post-common envelope binaries with circumbinary planets like those around NN\,Serpentis. 
We discuss a set of simple equations that determine upper and lower limits for orbital expansion and investigate the effect of initial eccentricity. 
We deduce the range of possible semi-major axes and initial eccentricity values of the planets prior to the common-envelope event. 
In addition to spherically-symmetric mass-ejection events, we consider planetary dynamics under the influence of an expanding disk. 
In order to have survived, we suggest that the present planets in NN\,Ser must have had semi-major axes $\simge 10~\AU$ and high eccentricity values which is in conflict with current observations. 
Consequently, we argue that these planets were not formed together with their hosting stellar system, but rather originated from the fraction of matter of the envelope that remained bound to the binary. 
According to the cooling age of the white dwarf primary of $10^6~\yr$, the planets around NN\,Ser might be the youngest known so far and open up a wide range of further study of second generation planet formation.
}  


\begin{document}

\maketitle

\section{Introduction}
\label{sec:intro}

The last two decades have seen a revolution in our knowledge of the properties and types of extrasolar planetary systems. 
Seemingly exotic objetcs like hot jupiters, super earths, pulsar planets, and circumbinary planets have dramatically expanded our understanding of planet formation.
A particular challange is the origin of circumbinary planets in post-common envelope binary (PCEB) systems \citep{Perets2010}: the predecessor systems are moderately close binaries with semi-major axes in the range of $\sim 1~\AU$ which, like Kepler 47 \citep{Orosz2012}, can contain circumbinary planets.

In PCEB systems, the common-envelope phase caused by the entry of the secondary star in the red giant envelope of the evolved primary redistributed the angular momentum budget of the stellar system and resulted in the formation of a close binary with orbital periods of just hours or days. Most of the primary's former mass is ejected because of the transfer of angular momentum and energy to the envelope on a time scale of weeks or months \citep{Ivanova2012}.\\

To date, a handfull of PCEB systems have provided evidence for the presence of planets, despite their dramatic dynamical evolutionary paths \citep{Horner2013}.
At the moment, the surest case is NN\,Ser, whose eclipse-timing variations can be explained by the presence of a planetary system consisting of two massive Jovian planets with roughly $7$ and $2\,\Mjup$, orbiting their hosting binary at distances of $3~\AU$ and $5~\AU$, respectively \citep[][]{Beuermann2010}.\\

The dynamical stability of the first fitted planetary configurations \citep{Beuermann2010} were investigated with detailed N-body simulations by \citet{Funk2011} and \citet{Horner2012} who found that some were unstable, whereas others were dynamically feasible. In the first fit, the outer planet's eccentricity was set to zero. \citet{Horner2012} did another analysis of the observational data but also varied the outer planet's eccentricity and obtained a new set of planetary parameters. In their study of the system's stability, most of their fits turned out to be unstable after short periods of time which initiated some controversary about the system's actual architecture. Recent dynamical analysis of other candidates such as HW Virginis \citep{Horner2012b} or HU Aquarii \citep{Horner2012c} also found high orbital instability in the proposed systems. Except for pure orbital fitting and stability analysis, intrinsic effects of the binary have been investigated and might also explain the observed transit timing variations \citep[see, e.g.,][]{Horner2012b, Horner2012c}.\\

Recently, \citet{Beuermann2013} used new observations and an equally extensive dynamical study to show that the family of stable solutions can now be constrained to be around a 2:1 resonance and moderate eccentricities (0.22 and 0.14 for the inner and outer planets, respectively)\footnote{Because the naming nomenclature for exoplanets has not yet been officially determined and several different nomenclatures appear in the literature, we use the names ``outer'' for the planet first detected and ``inner'' for the second planet detected; following the nomenclature proposed by \citet{Hessmann2010}, these would be equivalent to NN\,Ser (AB)b and NN\,Ser (AB)c, respectively.}. We summarize the results of \citet{Beuermann2013} in Table~\ref{tab:beuermann}.\\

\begin{table}[h]
\caption{Proposed NN\,Serpentis system configuration
  by \citet{Beuermann2013}.}
\begin{center}
\begin{tabular}{|c|c|c|c|}
\hline
\textbf{Name} & \textbf{Mass} & \textbf{a [AU]} & \textbf{e}\\
\hline
NN\,Ser A 			& 	$0.535 \Msol$ 	&	 - 		& 	-	 	\\
NN\,Ser B 			& 	$0.111 \Msol$ 	&	0.0043	& 	?		\\
NN\,Ser AB b (outer)	&	$6.97 \Mjup$	& 	5.387	&	0.144	\\
NN\,Ser AB c (inner)	& 	$1.73 \Mjup$	&	3.360	&	0.222 	\\
\hline
\end{tabular}
\label{tab:beuermann}
\end{center}
\end{table}

Even if the orbital structure of the planetary system in NN\,Ser has been determined, the question of its origin in such an exotic binary system is still unsolved.
Other than the very unlikely case of planetary capture, two distinct scenarios are possible.
\begin{itemize}
\item The planets formed jointly with the binary and survived the common-envelope event (First Generation Formation). 
\item The planets formed after the common-envelope event. Possible former planets did not survive (Second Generation Formation).
\end{itemize}
The second case is particularly fascinating as it implies a second phase of planet formation initiated by the common-envelope event. 
Recent studies conducted by \citet{Kashi2011} have shown that a fraction of 1 to 10 per cent of the envelope's mass can stay bound to the binary, falls back on it, flattens and forms a circumbinary disc out of which planets can easily form again.
If a first generation origin can be ruled out in the NN\,Ser system, this mechanism would have to be responsible for the present planets, opening up a whole new perspective on extrasolar planets.

\section{Simple estimates}
\label{sec:analytic}

We consider a planet with mass $m_{p}$ in a orbit with initial period $P_i$, velocity $v_i$, eccentricity $e_i$ and semi-major axis $a_i$. In addition, we neglect all kinds of dissipative effects on the system's energy and angular momentum budget. As a first approximation, stellar mass-ejection events can be modeled as a simple variation in the gravitational potential of the star, which affects the orbital parameters of any accompanying body in a way that it causes radial drift, changes in eccentricity, or an escape. A similar analysis was carried out by various authors. We refer the reader to \citet{Blaauw1961}, \citet{Alexander1976} or \citet{Perets2010}.

\begin{table}[h]
\caption{Analytic results for the final orbital parameters (f) of a planet on an iinitially circular orbit (i) for a given mass-loss factor $ \mu = M_f/M_i$.}
\begin{center}
\begin{tabular}{|c|c|c|}
\hline
\textsc{} & \textsc{Adiabatic} & \textsc{Instantaneous}\\
\hline
\hline
$a_f$ 		& 	$r_i \mu^{-1}$ 	&	$r_i / (2 - \mu ^{-1})$	 \\
$e_f$ 		& 	$0$ 				&	$\mu^{-1} - 1$	\\
$r_{peri}$ 	&	$r_i$	 		& 	$r_i$	\\
$r_{ap}$ 	& 	$r_i$ 			& 	$r_i/(2 \mu - 1)$ 	\\
$ E_f/E_i$ & $ \mu^{2}$ & $\mu$ \\
\hline
\end{tabular}
\label{tab:analytic}
\end{center}
\end{table}

\subsection{Initially circular orbits}

Two analytic limiting cases for stellar mass ejection and its effect on the orbiting planet(s) exist, characterized by the timescale of mass loss $\tau$ and the orbital period of the planet:

\begin{itemize}
\item $\tau \gg P_i$ (adiabatic mass loss) 
\item $\tau \ll P_i$ (instantaneous mass loss)
\end{itemize}

\noindent For both asymptotic regimes, we have summarized the analytic results in Table~\ref{tab:analytic}. In the case of instantaneous mass ejection, a mass-loss factor $\mu \le 1/2$ causes all of the planets to reach escape velocity and get lost, while for adiabatic mass loss no such boundary exists.

What conclusions can be drawn from this for NN\,Ser? The former binary system itself had a separation of about $1.44~\au$ \citep[][]{Beuermann2010}. \citet{Horner2012} derived a simple stability criterion from extensive numerical simulations, enabling us to estimate a lower limit for stable planetary orbits in the former binary. The main result of the stability criterion is that configurations with close encounters lower than three Hill radii (which is $\sim\!1.1~\AU$ for the secondary) are dynamically not feasible and require the planets to have semi-major axes of at least $\sim\!2.5~\AU$. \citet{Beuermann2010} argue that planets in the former NN\,Ser system should even have had semi-major axes of at least $3.5~\AU$, which is bigger than the typical extent of a $2~\Msol$ red giant star and the evolving CE. These results indicate that the former planets were not engulfed in the evolving Common Envelope. In addition, the mass of the envelope was much bigger than the total mass of both stellar cores, indicating that the planets did non suffer any significant drift as a result of the orbital shrinking of the binary. Hence, the planets can be approximately treated as decoupled from the binary evolution.\\

According to \citet{Beuermann2010}, we have $\mu\!\approx\!0.3$, implying that the original planets would have been lost if mass loss took place on an instantaneous timescale.
For adiabatic mass loss and assuming $e_i\!\approx\!0$ for both planets, the planets' original semi-major axes should be expanded by a factor of $\sim\!3.2$, implying $a_i < 0.3 \cdot a_f$.
Therefore, neglecting any mutual gravitational perturbations, the present planets may have orbited the original binary at separations of $a < 2~\AU$ if other interactions between the planet and the CE can be ruled out as explained in this section. Such close orbits are not dynamically feasible in the former binary.

On the other hand, considering an expansion factor of $\sim\!3.2$ for initially circular orbits and a lower limit of $\sim\!3.5~\AU$ for stable orbits, we conclude that first generation planetary candidates should now be detected at distances of more than $\sim\!10~\au$, which is obviously greater than the present planets' proposed semi-major axes.

\subsection{Initially elliptical orbits}

The orbital parameters of a planet are conserved if mass loss occurs adiabatically. 
However, for instantaneous mass-ejection they become time-dependant. 
For a given orbital configuration with semi-major axis $a_i$, eccentricity $e_i$, and a planet located at a distance $r_i$ at a polar angle $\phi_i$ at the moment of mass-ejection, the planet's new eccentricity and semi-major axis are given by
\begin{equation}
e_f = \frac{1}{\mu} \left(2 - \frac{1-e_i^2}{1+e_i \cos \phi_i} \right) - 1 ~~~ ; ~~~ a_f = \frac{a_i}{\mu} \cdot \frac{1-e_i^2}{1-e_f^2} ~~~ .
\end{equation}
Consequently, a particular mass-ejection scenario does result in planetary ejection if
\begin{equation}
\label{cond}
\frac{1}{\mu} \cdot \left( 2 - \frac{1-e_i^2}{1+e_i \cos \phi_i} \right) =  f(\mu , e_i, \phi_i) > 1 ~~~ .
\end{equation}
Relation~\ref{cond} can be directly translated into an escape probability $p_{esc} (\mu , e_i)$ defined as the fraction of unstable systems and stable configurations for a given mass-loss factor and given eccentricity by evaluating $f(\mu , e_i, \phi_i)$ for different polar angles $\phi_i$ (cf. Fig.~\ref{fig:esc-probability}).\\

In contrast to planets in circular orbits, a planet in an initially elliptical orbit can stay bound to the star even for $\mu < 0.5$ if the mass-ejection event happens near apastron, i.e., the planet moves in a sufficiently eccentrical orbit. 
On the other hand, the planet can escape from the system if it happens to be located near periastron even for $\mu > 0.5$. 
Another consequence is that the orbit can be circularized for particular configurations: if the mass-ejection event takes place near the apastron, an initial eccentricity of $e_i = 1-\mu$ will result in a circular orbit and an orbital expansion factor of $a_f / a_i = 2 - \mu$.\\ 

What are the implications for NN\,Ser? For planets at a distance of a few AU we have orbital periods of a few years and consequently $\tau \ll P_i$ for a typical CE event \citep{Ivanova2012}. Again, we can neglect drag effects or dissipative interactions between the envelope and the planets as well as orbital drift due to the evolution of the binary, as explained in the last section. In the case of instantaneous mass loss, from Fig.~\ref{fig:esc-probability} it becomes clear that a mass-loss factor as dramatic as $\mu \approx 0.3$ requires the planets to have very high initial eccentricities of $e_i \approx 0.4$ individually to ensure at least small survival probabilities. However, high original eccentricities for {\it both} planets makes it even more difficult to obtain stable first-generation solutions for both planets.

\bef
\resizebox{\hsize}{!}{\includegraphics{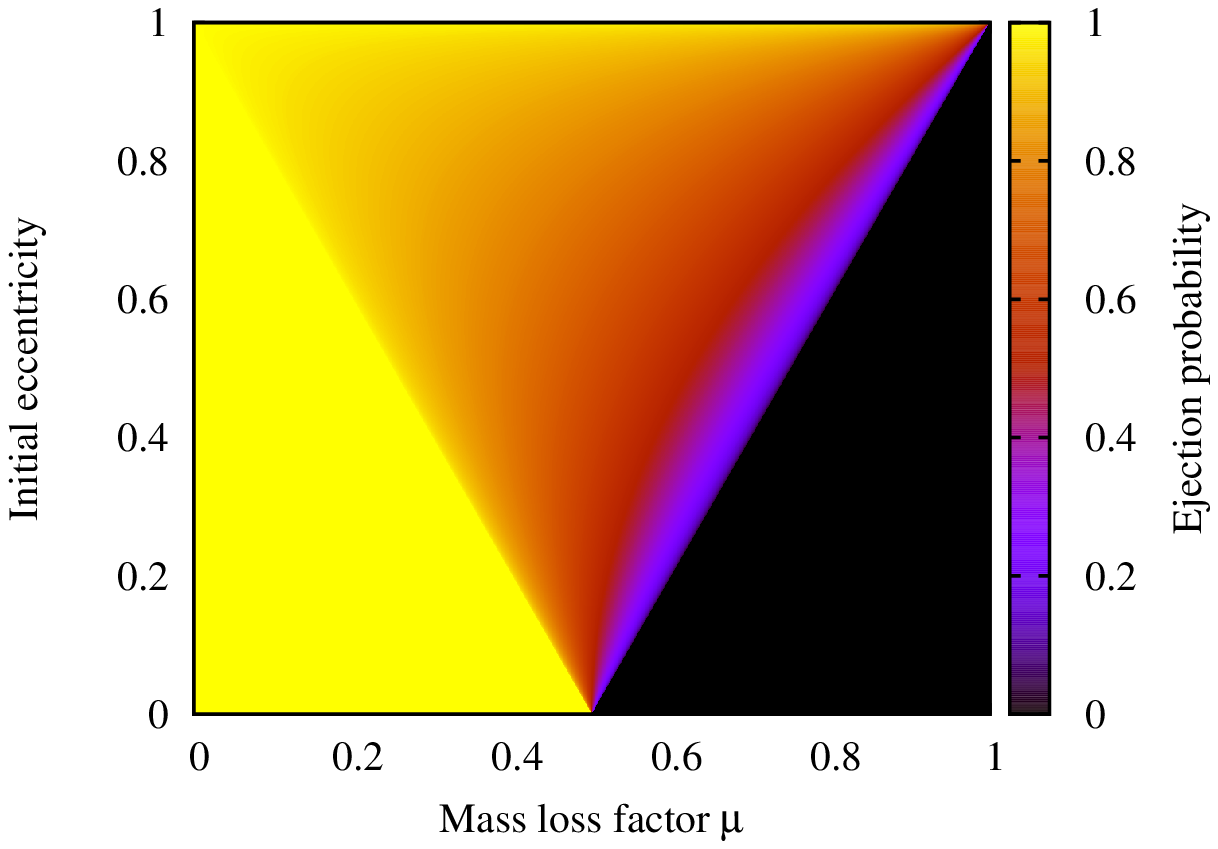}}
\caption{Escape probability for a planet with varying initial eccentricity for different mass-loss factors in an instantaneous mass-ejection case. A mass-loss factor of $\approx 0.3$ allows the planet to escape the system if initial eccentricity is lower than $\approx 0.4$.}
\label{fig:esc-probability}
\eef 

\section{Gas drag estimates}

So far, we have derived analytic results of the orbital parameters for the two asymptotic regimes of slowly and rapidly changing gravitational potentials without any friction effects due to gas drag. To model the gas ejection event and estimate the effect of gas drag, we chose a self-consistent, exponential density profile 
\begin{equation}
\rho(r,t) = \begin{cases} 
\rho_0  \exp{[\beta(r_{gas}(t)) \cdot r]} & 0 \leq r \leq r_{gas}(t) \\
0 & r > r_{gas}(t)
\end{cases}
\end{equation}  
($\beta > 0$) and a linear velocity profile characterised by the outer rim velocity $v_{max}$ and the initial outer boundary $r_{max,0}$ of the expanding envelope, mainly based on results obtained by \citet{Taam2010} and \citet{Ricker2012} and the Sedov profile for blast wave propagation. 

For a given mass, $\beta(r_{gas})$ is determined by the total amount of gas $M_{gas}$ via
\begin{equation}
M_{gas} = 4 \pi \rho_0 \int\limits_0^{r_{gas}(t)} r^2 \exp{(\beta \cdot r)} \D r ~~~ .
\end{equation}

Gas drag effects \citep[see, e.g.,][]{Kokubo2000} result in the additional force
\begin{equation}
\vec{F}_{d} = \underbrace{\frac{1}{2} \pi R_{planet}^2 c_d }_{:=c_{p}} \cdot \rho(r_p) \cdot |\vec{u}| \cdot \vec{u} ~~~ ,
\end{equation}
where $\vec{u} = \vec{v}_{gas} - \vec{v}_{planet}$. 
A quick upper limit for $ | \vec{F}_d | / | \vec{F}_g | $ can be obtained by considering the planet in an initially circular orbit and also assuming $\vec{v}_p \perp \vec{v}_{gas}$, as well as $|\vec{v}_p| = \sqrt{G M / r_p}$ and $|\vec{u}|^2 = |\vec{v}_p|^2 + |\vec{v}_{max}|^2$:
\begin{equation}
\frac{|\vec{F}_d|}{|\vec{F}_g|} = \frac{c_p \rho(r) [\frac{GM}{r} + |\vec{v}_{max}|^2 ]}{\frac{G M m_p}{r^2}} = 
\rho_0 \frac{c_p}{m_p} e^{\gamma} r_p \left[1 + \frac{|\vec{v}_{max}|^2}{|\vec{v}_p|^2} \right] ~~~ .
\end{equation}
For an initially close planet ($r_p \approx 2~\au $), we have $|\vec{v}_{gas}|^2 \ll |\vec{v}_p|^2$ and $\beta \approx 10$. 
Using $m_p \approx 5~\Mjup$ for a massive Jovian planet, $\rho_0 = 10^{-15}~\g / \cm^3$ for the gas sphere and typical values for the specific drag coefficient like $c_p \approx 10^{12}~\cm^2$ \citep{Kokubo2000} leads to $|\vec{F}_d| / |\vec{F}_g| \approx 10^{-4}$.
For a distant planet ($r_p \approx 20~\au $) we have $\gamma \rightarrow 0$ and can set $|\vec{u}| \approx |\vec{v}_{max}| \approx 10~\km / \s $ so that $|\vec{F}_d| / |\vec{F}_g| \approx 10^{-10}$.
These results imply that when there is spherical symmetry geometric gas drag effects are not expected to cause any significant deviations from our solutions even in very massive ejection events. 

For the density profile discussed in this section, we also conducted a number of simulation runs with different gas velocities and $\mu = 0.3$ where we placed a planet in an initially circular orbit between $2.0~\au$ and $7.0~\au$. For $v_{max} > 10~\km / \s$, virtually all planets have already exceeded the escape velocity.
For $1~\km / \s < v_{max} < 10~\km / \s$, the planets' semi-major axes expanded at least by a factor of $\sim 3$. 
As expected, the planets ended up with significantly eccentric orbits.\\

Based on these results, we conclude that it appears to be impossible to produce circumbinary planets resembling those in NN\,Ser using a spherically symmetric density profile if the planets were not part of the CE and did not interact with it. The envelope densities and hence geometric drag forces of the expanding envelope which might have resulted in the retention of the original planets are too small to cause any significant inward directed drift.

\section{Disc-shaped mass-ejection}

Common-envelope ejection events are, admittedly, only roughly spherical-symmetric, since the envelope is given the considerable angular momentum of the secondary star.
In fact, hydrodynamical simulations of common-envelope events \citep[see e.g.,][]{Ivanova2012, Taam2010, Ricker2012} showed a significant amount of mass concentration in the orbital plane of the binary star. 
Thus, we conducted a few simulations of planetary dynamics under the influence of a thin expanding disc (see \citet{Hure2011} for further information as well as a far more rigid approach).\\
  
For $z_{gas} \ll r_{gas}$, $\vec{r}(r,z,\phi) \approx \vec{r}(r,\phi)$ and the gravitational potential of the disc can be approximated as
\begin{equation}
U( \vec{r} ) \approx -2 G z_{gas} \cdot \int\limits_{0}^{r_{gas}} \int\limits_{0}^{2\pi} \frac{\rho (\vec{r'})}{|\vec{r}-\vec{r'}|} r' \D r' \D \phi' ~~~ .
\end{equation}  
This integral can be easily approximated by that of a disc split into a number of annuli segments ($\Delta r$,$\Delta \phi$), where a particular segment's mean density and position vector is evaluated in its middle.
The gravitational potentials of thin exponential discs with $M = 1~\Msol$, $\rho_0 = 10^{-15}~\g \pcc$, $z_{gas} = 0.05~\au$,  and different radii are plotted in Fig.~\ref{fig:disc-potential}. All disc potentials feature a prominent potential minimum located at the disc's outer edge whose depth is predominantly determined by the disc's extension.\\

Because of the substantially lower volume of the disc, gas densities are expected to be significantly higher and gas drag effects might be an issue. The density profile parameter $\beta$ drops more slowly and the outer rim gas densities are higher. 
Estimating the effect of friction is complicated, as no analytic result exists for the potential of a finite disc with non-constant density profile. For a moderately thin disc ($h\approx r_{max}/10$) with two $\AU$ in radius and an exponential density profile with $\rho _0 \sim 10^{-15}~\g \pcc$, numerical results give an outer rim density of $~3 \cdot 10^{-6} ~\g \pcc$, which is only one or two orders of magnitude higher than the outer rim density of a comparable gas sphere. Consequently, geometric gas drag effects are still negligible.\\

When this model is applied to NN\,Ser, an initially close planet ($a < 5~\au$) starts to drift radially inwards onto the expanding disc as the potential minimum approaches. Entering the disc, the planet is exposed to gas densities in the range of $\approx 10^{-6}~\g \pcc$. While the disc continues its expansion, the planet slows down because of potential effects (see Fig~\ref{fig:disc-potential}) and suffers from outwards directed gravitational forces inside the disc. 
As a consequence, the planet increases its radial distance from the disc's center and is carried towards the disc's edge by gravitational forces again. In fact, most of our simulations of planetary dynamics under the influence of a massive expanding disc resulted in the ejection of the planet. In this approach, cylindrical symmetry helps the planets to get unbound.

\bef
\resizebox{\hsize}{!}{\includegraphics{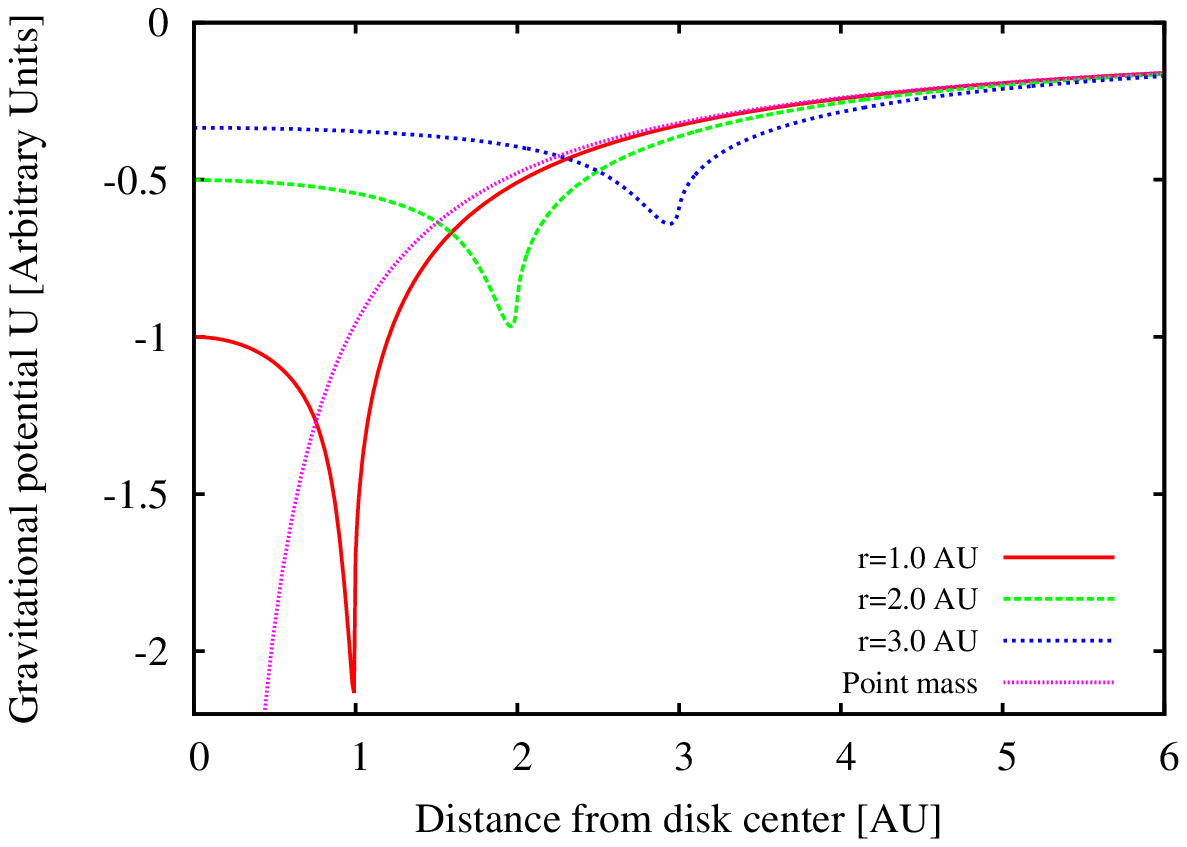}}
\caption{Gravitational potential for a thin disc with $M = 1~\Msol$, $\rho_0 = 10^{-15}~\g / cm^3$, $z_{gas} = 0.05~\au$, and varying disc radii in the disc plane. Potential values are normalized to $U(0)$ for $r=1.0~\au$. A point-mass potential is plotted for comparison. The disc potential is characterized by a prominent potential minimum located at the disc's rim especially in compact discs.}
\label{fig:disc-potential}
\eef

\section{Conclusions and discussion}
\label{sec:conclusions}

We analyzed the general impact of mass loss in both circular and eccentrical orbits. In NN\,Ser, simple geometric arguments and semi-analytic stability criteria suggest that dissipative interactions between the common envelope and the planets as well as orbital variations due to the evolution of the hosting binary can be neglected. Rather, our analytic work implies that the dramatic mass loss that led to the formation of the close binary system NN\,Ser should have resulted in a significant orbital expansion if not loss of the planets, especially for initially eccentric orbits.
As a consequence, a first generation scenario demands that the original planets had highly elliptical orbits and that the present planets should have semi-major axes of more than $10~\AU$. 
The latter is clearly in conflict with the current planets' inferred orbital parameters. 
Furthermore, we examined the effect of cylindrical symmetry on the planetary dynamics and found that this scenario also helps the planets' ejection.\\

Considering all of these effects, we argue that the planets around NN\,Ser could not have been formed together with the hosting binary star. Hence, these planets should have formed out of the back-falling material of the CE \citep[e.g.,][]{Kashi2011}, i.e. they should be second generation planets \citep{Perets2010}.
Given a cooling age of about $10^6~\yr$ \citep[see, e.g.,][]{Beuermann2010} these planets are then the youngest planets known so far. 
Further efforts are essential in order to constrain orbital parameters of the planets, to get a better idea of the common envelope ejection phase, and to estimate possible planet-envelope interactions and ascertain the timescale for mean-motion resonance locking in this particular system. 
In any case, NN\,Ser should cast much light on this fascinating form of planetary evolution. 

\begin{acknowledgements}
We would like to thank Stefan Dreizler for valuable discussions. We would also like to thank the anonymous referee for useful advice. We gratefully acknowledge funding from the Deutsche Forschungsgemeinschaft via Research Training Group (GrK) 1351 \textit{Extrasolar Planets and their Host Stars}.
\end{acknowledgements}

\bibliography{astro-marcel.bib}
\bibliographystyle{aa}

\end{document}